# Hot-Ham: an accurate and efficient E(3)-equivariant machine-learning electronic structures calculation framework

Zhixin Liang[1, †], Yunlong Wang[1, †], Chi Ding[1], Junjie Wang[1,*], Hui-Tian Wang[1], Dingyu Xing[1], and Jian Sun[1,*]

[1]*National Laboratory of Solid State Microstructures, School of Physics and Collaborative Innovation Center of Advanced Microstructures, Nanjing University, Nanjing, 210093, China*

## Abstract

The combinations of machine learning with ab initio methods have attracted much attention for their potential to resolve the accuracy-efficiency dilemma and facilitate calculations for large-scale systems. Recently, equivariant message passing neural networks (MPNNs) that explicitly incorporate symmetry constraints have demonstrated promise for interatomic potential and density functional theory (DFT) Hamiltonian predictions. However, the high-order tensors used to represent node and edge information are coupled through the Clebsch-Gordan tensor product (CGTP), leading to steep increases in computational complexity and seriously hindering the performance of equivariant MPNNs. Here, we develop High-order Tensor machine-learning Hamiltonian (Hot-Ham), an E(3) equivariant MPNN framework that combines two advanced technologies local coordinate transformation and Gaunt tensor product (GTP) to efficiently model DFT Hamiltonians. These two innovations significantly reduce the complexity of tensor products from $O(L^6)$ to $O(L^3)$ or $O(L^2 log^2 L)$ for the max tensor order *L*, and enhance the performance of MPNNs. Benchmarks on several public datasets demonstrate its state-of-the-art accuracy with relatively few parameters, and the applications to multilayer twisted moiré systems, heterostructures and allotropes showcase its generalization ability and high efficiency. Our Hot-Ham method provides a new perspective for developing efficient equivariant neural networks and would be a promising approach for investigating the electronic properties of large-scale materials systems.

---

[†] These authors contributed equally to this work.
[*] Corresponding author. J.S (jiansun@nju.edu.cn); J.W. (wangjunjie@nju.edu.cn)

# Introduction

Density Functional Theory (DFT) has become one of the most popular tools for studying the structure, properties, and reactions of materials at the atomic level. While DFT provides a quantum mechanical framework for electronic structure calculations, solving the Kohn-Sham equations requires significant computational resources, restricting its application to systems with a limited number of atoms. In contrast, empirical tight-binding such as Slater-Koster method[1] can typically be evaluated orders of magnitude faster than first-principles methods due to their simple mathematical form, hence enabling large-scale electronic calculations. However, empirical tight-binding often comes at the cost of reduced accuracy compared to DFT methods and limited transferability across different systems.

Emerging machine learning (ML) techniques have been increasingly applied to model electronic structures[2–14], showing great promise in providing a way to address this accuracy-efficiency dilemma. Among the various approaches, graph neutral networks[15–18] (GNNs) have become the dominant choice to characterize the graph structures. Schütt et al.[5] proposed a GNN model SchNorb to predict Hamiltonian as a linear combination of spherical harmonics, however, it does not incorporate the symmetry as a priori knowledge. An important property of Hamiltonian is its equivariant transformation under the Euclidean group on the three-dimension space (E(3) group), which contains translations, rotations, and inversion. Models that explicitly incorporate equivariance constraints are more data-efficient and generalizable, making them promising for complex prediction tasks. PhiSNet[6], QHNET[7], and ACEHamiltonians[8] achieve physical correctness under translation and rotation transformations, but all of them would be struggling with periodic materials due to a lack of parity symmetry[12]. The N-center representation equivariant framework proposed by Nigam[9] achieved E(3)-equivariant Hamiltonian using Gaussian process regression (GPR). However, GPR usually has limited generalization ability and is more computationally expensive compared with GNNs such as message-passing neural networks[19] (MPNNs). Combining E(3) equivariance constraints with MPNNs, Gong et al[11] and Zhong et al[12] proposed DeepH-E3 and HamGNN respectively and showcased excellent accuracy in Hamiltonian predictions across diverse materials.

Despite the remarkable success of E(3)-equivariant neural networks (E(3)-ENNs), these models face significant limitations due to the high computational complexity of equivariant operations that employ Clebsch-Gordan tensor product (CGTP) of irreducible representations (irreps). The CGTP couples tensors with orders up to $L$ to produce new features, leading to a computational complexity of $O(L^6)$ for the full operation. This steeply increasing complexity hinders the application of E(3)-ENNs for predicting physical quantities that require high-order tensors. To address this issue, Passaro and Zitnick[20] introduced local coordinate transformation, simplifying CGTP to SO(2) convolution thus reducing the complexity to $O(L^3)$. Recent E(3)-equivariant Hamiltonian frameworks DeepH-2[13] and DeePTB[14] have adopted this approach and show state-of-the-art accuracy in Hamiltonian predictions. However, SO(2) convolution is only used for equivariant convolutions, a special case of CGTPs. As there are no specific local coordinate transformations for arbitrary equivariant features, this method is not suitable for equivariant feature interactions (e.g. HamGNN) and equivariant many-body interactions (e.g. MACE[21]). Xin et al[22] proposed another fast and accurate approach for spherical harmonics products, and Luo et al[23] also proposed a similar tensor product named Gaunt tensor products (GTPs), which are more efficient and general as compared with SO(2) convolution. Nonetheless, since GTPs inherently exclude antisymmetric parity, they are hardly utilized in E(3)-ENNs, especially for the Hamiltonian predictions where antisymmetric tensors are indispensable.

In this work, we propose High-order Tensor machine learning Hamiltonian (Hot-Ham), an E(3)-equivariant message passing neural network framework designed for efficient Hamiltonian representation. By combining local coordinate transformations and GTPs, our framework not only effectively leverages their strengths but also compensates for their limitations, reducing tensor products from $O(L^6)$ to $O(L^3)$ or $O(L^2 log^2 L)$ and establishing an efficient E(3)-ENN for electronic structure calculation.

We benchmark Hot-Ham on several public datasets and compare its performance with other Hamiltonian models. Our results show that Hot-Ham achieves superior performance, while maintaining significantly smaller parameter size (within 2M) compared to most models. To demonstrate the generalization ability and efficiency, we train Hot-Ham on non-twisted $MoS_2$, graphene/hexagonal boron nitride (h-BN) heterostructures and phosphorus allotropes, and test it on various structures out of datasets. The accuracy shows that our model is able to accurately predict electronic structure even though the crystal structure is not included in the training set, with computational costs much lower than DFT. The accuracy, generalization, and efficiency of Hot-Ham significantly mitigate the accuracy-efficiency trade-off dilemma of DFT, opening up new possibilities for large-scale electronic structure calculations and novel functional material discovery.

# Results

## Equivariant Hamiltonian

The properties of physical systems generally exhibit symmetries under transformations of a set of atoms, leading to the concept of equivariance naturally. Formally, a function $f: X \to Y$ is equivariant for $X$ and $Y$ with respect to group $G$ if $f(g_X x) = g_Y f(x)$ for $\forall g \in G$, where $g_X$ and $g_Y$ are group representations on $X$ and $Y$. ENNs are guaranteed to preserve equivariance under a change of coordinates because they are composed of equivariant functions.

In GNNs, the structure is represented as a graph, where each node corresponds to an atom, and each edge represents a connection between atoms within a cutoff radius $r_c$. Given a set of atoms, the two key problems of designing ENNs are how to encode the atom positions $\{\mathbf{r}_i\}$ and chemical elements $\{\mathbf{Z}_i\}$ as equivariant features for nodes and edges, and how to relate these features to equivariant Hamiltonians.

Traditionally, DFT Hamiltonians can be calculated on various bases, such as plane-wave basis[24–26], linear combination of atomic orbitals (LCAO) basis[27–30], and so on. Specifically, the LCAO basis has the form $\phi_{nlm}(\mathbf{r} - \mathbf{r}_i) = R_{nl}(|\mathbf{r} - \mathbf{r}_i|)Y_{lm}(\mathbf{r} - \mathbf{r}_i)$, where $i$ is the atom index, $R_{nl}$ and $Y_{lm}$ are radial function and spherical harmonics

respectively with projectors' multiplicity $n$, angular momentum quantum number $l$ and magnetic quantum number $m$. On the one hand, this form restricts the interaction calculations within a cutoff distance, which is consistent with the finite receptive field of ENNs. On the other hand, spherical harmonics, which is also refered to as spherical tensors, satisfy rotation and inversion equivariance (O(3) group), and have been widely utilized to represent equivariant features inside ENNs[21,31–33]. Therefore, among various kinds of bases, LCAO-basis Hamiltonians are suitable learning targets for ML Hamiltonian model. According to Wigner-Eckart theorem, the LCAO-basis Hamiltonian element $H_{ij,n_1l_1m_1n_2l_2m_2} = \langle i, n_1l_1m_1|\hat{H}|j, n_2l_2m_2\rangle$ between $i,j$ atoms can be decomposed into several learnable spherical tensors $x_{ij,c_3l_3m_3p_3}$:

$$\langle i, n_1l_1m_1|\hat{H}|j, n_2l_2m_2\rangle = \sum_{l_3=|l_1-l_2|}^{l_1+l_2} \sum_{m_3=-l_3}^{l_3} C_{l_1m_1l_2m_2}^{l_3m_3} x_{ij,c_3l_3m_3p_3} \quad (1)$$

Here $c_3$ is the channel index that is determined by the $n$ and $l$ of two atomic orbitals, and $p_3 \in [1,-1]$ is the parity index. And $C_{l_1m_1l_2m_2}^{l_3m_3}$ is the Clebsch-Gordan coefficient. The atomic orbitals basis for $i$th atom $|i, n_1l_1m_1\rangle = \phi_{n_1l_1m_1}(\mathbf{r} - \mathbf{r}_i)$ has a parity of $p_1 = (-1)^{l_1}$. So the parities of $H_{ij,n_1l_1m_1n_2l_2m_2}$, and thus $x_{ij,c_3l_3m_3p_3}$, are determined by the two atomic orbitals basis as $p_3 = (-1)^{l_1+l_2}$. $x_{ij,c_3l_3m_3p_3}$ represents equivariant feature of atom for onsite block if $i = j$, or edge for hopping block if $i \neq j$. Hence, the aim of ENNs is to predict high-order tensor features that comprise a direct sum of irreps of the O(3) for all nodes and edges.

Chemical elements and interatomic distance are invariant physical quantities and are used to construct 0-order tensors, i.e., scalars. The relative positions of atoms are mapped into spherical harmonics to serve as high-order tensors, which automatically ensure the invariance of ENNs under translation. After a series of equivariant operations, the final equivariant features are used to reconstruct Hamiltonians in a block-wise manner through (1). It should be noted that the equivariant operations have to be carefully designed, not only to preserve E(3) equivariance but also to achieve high efficiency for the neural networks. One of the main equivariant operations is the E3linear transformations. The linear operations mix the features between channels in the same irreps. For the common feature representation $X^L = \oplus_{l=0}^L x_l$, which represents the direct sum of spherical tensors $x_l$ up to order $L$, the linear transformations can be conducted in a complexity of $O(L^2)$. The other main equivariant operations are tensor products. These operations couple tensors $x_1$ and $x_2$ in different irreps through the Clebsch-Gordan coefficient, and yield new tensors $x_3$ that is E(3)-equivariant:

$$x_{3,l_3 m_3} = (x_1 \otimes_{CG} x_2)_{l_3 m_3} = \sum_{m_1 m_2} C^{l_3 m_3}_{l_1 m_1 l_2 m_2} x_{1,l_1 m_1} x_{2,l_2 m_2} \qquad (2)$$

However, when applying (2) to features like $X^L$, the complete tensor products will lead to an $O(L^6)$ complexity, which seriously hinders the application of ENNs to Hamiltonian predictions, where high-order tensors are essential to represent atom orbitals and their interactions, thus demanding new approaches to achieve efficient tensor products

## Efficient tensor product operations

SO(2) convolution is one of the approaches to accelerating the tensor product operations. In the field of ENNs, filters can be defined as spherical harmonic acting on relative positions $\mathbf{r}_{ij}$. For convolution operation, i.e. tensor product between feature $x_{1,l_1}$ and filter $x_{2,l_2} = Y_{l_2}(\mathbf{r}_{ij})$, the complexity can be reduced through local coordinate transformation. By rotating the embeddings' primary axis to align with the edge vectors, the filter $Y_{l_2 m_2}(\mathbf{r}_{ij})$ becomes sparse: $Y_{l_2 m_2}(\mathbf{r}_{ij}) = 0$ if $m_2 \neq 0$. This eliminates the summation over $m_2$, reducing (2) to a 2D matrix multiplication. By rearranging the Clebsch-Gordan coefficients and relevant weights, the tensor products between rotated features like $X^L$ and filters with orders up to $L$ are simplified to $O(L)$ 2D matrix multiplications, which accounts for $O(L^3)$ cost. Additionally, an $O(L^2)$ operation is needed for each irrep among $X^L$ to rotate align with the edge vector or rotate back, accounting for two extra $O(L^3)$ cost.

GTP is another more efficient, and general approach. Gaunt coefficients[34] $G^{l_3 m_3}_{l_1 m_1 l_2 m_2}$ are defined as the integrals of three spherical harmonics products, and are related to Clebsch-Gordan coefficients via constant factors $\tilde{C}^{l_3}_{l_1 l_2}$ that are independent on magnetic quantum number: $G^{l_3 m_3}_{l_1 m_1 l_2 m_2} = \tilde{C}^{l_3}_{l_1 l_2} C^{l_3 m_3}_{l_1 m_1 l_2 m_2}$. Luo et al[23] propose new perspectives that GTPs calculate the coefficients of spherical functions based on spherical harmonics:

$$\sum_{l_1}^{L_1}\sum_{l_2}^{L_2}(x_{1,l_1}\otimes_{Gaunt}x_{2,l_2})_{l_3m_3}=\int_0^{2\pi}\int_0^{\pi}F_{L_1}(\theta,\psi)F_{L_2}(\theta,\psi)Y_{l_3m_3}(\theta,\psi)\sin\theta\,d\theta d\psi \quad (3)$$

for which $F_{L_i}(\theta,\psi)=\sum_{l=0}^{L_i}\sum_{m=-l}^{l}x_{i,lm}Y_{lm}(\theta,\psi)$ is a square-integrable spherical function, and appropriate base transformations will be helpful to parallel these operations. An illustration is presented in Fig. 1a. By changing the spherical harmonics bases into 2D Fourier bases, spherical function $F(\theta,\psi)=F_{L_1}(\theta,\psi)F_{L_2}(\theta,\psi)$ will be simplified into 2D convolution, which can be accelerated via 2D FFT in $O(L^2 logL)$. Finally, the coefficients of $F(\theta,\psi)$ are converted back to spherical harmonic representations. These base transformations account for $O(L^3)$ cost. And we call it GTP(2D-FB) in this paper to distinguish it from the method GTP(sphere-grid) proposed by Xie et al[35] through spherical convolution, as shown in Fig. 1b. Features are first projected on a sphere grid through an inverse FFT. Following an $O(L^2)$ element-wise producting, a FFT is performed to convert back to spherical harmonics. The two FFT accounts for an $O(L^2 log^2 L)$ complexity. Therefore, GTP(sphere-grid) becomes theoretically the fastest method among the four kinds of tensor products. Similarly, the coefficients of both GTP methods can be further sparse via local coordinate transformation in the case of convolution. However, it should be noted that there is a difference between GTP and CGTP, because $\tilde{C}_{l_1l_2}^{l_3}$ vanish when $l_1+l_2+l_3=2k+1$ non-negative integer $k$, thus leading to the lack of antisymmetric tensors in GTP. Nonetheless, GTPs still adhere to E(3) equivariance requirements.

We conduct experiments on the convolution operations using these four tensor products to demonstrate their efficiency, as shown in Fig. 1c. The origin CGTP implemented by the e3nn package is unexpectedly the most time-consuming, while GTPs, especially GTP(sphere-grid), achieve the highest efficiency. These results, as well as the fact that GTP is suitable for various tensor products not only for convolution, indicate that GTP is a promising operation to achieve high-efficiency equivariant MPNNs.

## Equivariant message passing neural network

The Hot-Ham model architecture is illustrated in Fig. 2. Atomic number $Z_i$,

interatomic distance $|\mathbf{r}_{ij}|$ and direction information $Y(\mathbf{r}_{ij})$ are embedded to generate initial node features $v_i^{(0)}$ and edge features $e_{ij}^{(0)}$. Features are structured as a direct sum of symmetric tensors $x_{c_0lm,(-1)^l}$ and antisymmetric tensors $x_{c_1lm,(-1)^{l+1}}$. Since only symmetric tensors can be generated by GTP, we need to introduce antisymmetric tensors at least once through CGTP. Given that including antisymmetric tensors in intermediate layers would result in about double computational costs, we choose to perform CGTP in the last layer through SO(2) convolution. Node and edge symmetric tensor features are encoded and aggregated to update iteratively in each GTP convolution layer, then are extended to include antisymmetric tensor features in the last SO(2) convolution layer. Finally, these features are transformed into Hamiltonian matrix block $H_{ij}$ through Wigner-Eckart theorem in the Readout layer.

**Feature initialization**

As shown in Fig. 2b, the edge features are defined as weighted spherical harmonics through a linear transformation, and node features are the mean of edge features:

$$e_{ij,lmc}^{(0)} = w_{ij,lc}(|\mathbf{r}_{ij}|)Y_{lm}(\mathbf{r}_{ij})$$
$$v_i^{(0)} = \frac{1}{|N(i)|}\sum_{j \in N(i)} e_{ij}^{(0)} \qquad (4)$$

Here $N(i)$ are neighbor atoms of atom $i$. $w_{ij,lc}$ are weights that contain information about the species of two connected atoms and their interatomic distance. Specifically, $Z_i$ and $Z_j$ are mapped by one-hot encoding and then learnable multi-layer perceptron (MLP) into two species vectors, and $|\mathbf{r}_{ij}|$ is expanded by radial bases functions (RBF) such as Gaussian, Bessel, and Chebyshev basis. Subsequently, these species vectors and distance expansions are concatenated and transformed by MLP into $w_{ij,lc}$.

**Convolution**

In convolution layers, local environment information including connected node features $v_i^{(t)}$ and edge feature $e_{ij}^{(t)}$ are used to encode edge messages through the tensor products between the weighted concatenations $C_{ij,cl_1m_1}^{(t)}$ of features and weighted filters $F_{ij,l_2m_2}^{(t)}$:

$$C_{ij,c_{out}l_1m_1}^{(t)} = w_{ij,c_{in}c_{out}l_1}^{1,(t)}(|\mathbf{r}_{ij}|) \cdot \left(v_i^{(t)} \| v_j^{(t)} \| e_{ij}^{(t)}\right)_{c_{in}l_1m_1}$$

$$F_{ij,l_2m_2}^{(t)} = w_{ij,l_2}^{2,(t)}(|\mathbf{r}_{ij}|) \cdot Y_{l_2m_2}(\mathbf{r}_{ij}) \quad (5)$$

$$m_{ij,c_{out}l_3m_3}^{(t)} = \sum_{l_1m_1l_2m_2} C_{l_1m_1l_2m_2}^{l_3m_3} C_{ij,c_{out}l_1m_1}^{(t)} F_{ij,l_2m_2}^{(t)}$$

Here $w_{ij,c_{in}c_{out}l}^{1,(t)}$ and $w_{ij,l_2}^{2,(t)}$ are distance-dependent weight functions, and $\|$ denotes concatenation. To introduce nonlinearity without breaking E(3) equivariance, a gate (activation function) must be carefully designed. In Hot-Ham, we use a gate like ref.[36], where scalar and pseudoscalar components are nonlinear by SiLU and tanh respectively, while others are scaled by learnable scalars that are nonlinear by sigmoid. Messages are then used to update edge and node features simultaneously. New edge features are updated with the residual net like architecture[32]:

$$\tilde{e}_{ij,clm}^{(t+1)} = w_{0,lc}^{(t)} e_{ij,clm}^{(t)} + w_{1,lc}^{(t)} \text{Gate}\left(m_{ij,clm}^{(t)}\right) \quad (6)$$

For node features, messages are first propagated to each node along edges and then aggregated by mean function. Similarly, the old node features are combined with the messages to form the new node features:

$$\tilde{v}_{i,clm}^{(t+1)} = w_{2,cl}^{(t)} v_{i,clm}^{(t)} + w_{3,clm}^{(t)} \left(\frac{1}{|N(i)|} \sum_{j \in N(i)} \text{Gate}\left(m_{ij,clm}^{(t)}\right)\right) \quad (7)$$

## Layer normalization

Normalization is a crucial technique in deep learning models used to stabilize and enhance training. Layer normalization is one of the most common methods applied in ENNs due to its independence of batch size, which is typically small in Hamiltonian model training. To preserve equivariance, feature components belonging to different irreps are normalized separately. For scalar features $\tilde{f}_{z,0}^{(t)}$, we use a layer normalization like ref.[11,31]:

$$f_{z,c00}^{(t)} = \gamma_{c0} \frac{\tilde{f}_{z,c00}^{(t)} - \mu_{00}}{\sigma_0 + \varepsilon} + \beta_{c0} \quad (8)$$

where $z$ represents atom index $i$ for $\tilde{v}_{i,clm}^{(t)}$ or edge index $ij$ for $\tilde{e}_{ij,clm}^{(t)}$, $\mu_{lm} =$

$\frac{1}{N_Z N_C} \sum_{Z=1,c=1}^{N_Z,N_C} \tilde{f}_{Z,clm}^{(t)}$ and $\sigma_l = \sqrt{\frac{1}{N_Z N_C} \sum_{Z=1,c=1}^{N_Z,N_C} \left| \tilde{f}_{Z,clm}^{(t)} - \mu_{lm} \right|^2}$ are mean and standard deviation of features with degree $l$. $\varepsilon$ is a small number used to avoid numerical instability when $\sigma_0$ is close to zero. $\gamma_{cl}$ and $\beta_{cl}$ are learnable parameters for affine transformation. While for pseudoscalar or tensor with $l > 0$, we simply subtract the mean:

$$f_{z,clm}^{(t)} = \tilde{f}_{z,clm}^{(t)} - \mu_{lm} \tag{9}$$

**Readout**

The Hamiltonian matrix block $H_{ij,l_1 l_2}$ between atom $i$ with orbital angular momentum $l_1$ and atom $j$ with orbital angular momentum $l_2$ takes the form of a direct product: $l_1 \otimes l_2$, a $(2l_1 + 1) \times (2l_2 + 1)$ matrix with parity $p = (-1)^{l_1 + l_2}$. After several layers of convolution iteration, the final features have to be passed through a linear layer to rearrange into a form $\bigoplus_{l=|l_1-l_2|}^{l_1+l_2} x_{ij,lp}$, and then transform to $H_{ij,l_1 l_2}$ through (1). Incorporating the prior Hermitian property for Hamiltonian will be beneficial to reducing loss. To enforce the Hamiltonian to be Hermitian, we define the final Hamiltonian matrix block as: $\tilde{H}_{ij} = \frac{1}{2}(H_{ij} + H_{ji}^*)$.

## Performance of Hot-Ham

To benchmark the accuracy of our models, we apply Hot-Ham to public datasets[11], including monolayer graphene, monolayer MoS$_2$, and bilayer graphene, as reported in ref. [10]. For each dataset, we train two models using GTP(2D-FB) and GTP(sphere-grid) respectively in the GTP convolution layers and compare the results with other models. To further demonstrate the generalization and efficiency of our approach, we also apply Hot-Ham to study the electronic structure of multilayer twisted MoS$_2$, graphene/h-BN heterostructures, and phosphorus allotropes.

## Benchmark accuracy

We compare the mean absolute error (MAE) of Hamiltonian matrix elements and the number of training parameters for Hot-Ham with other models, including DeepH-E3, HamGNN, DeepH-2, and DeePTB. The results for the monolayer graphene and monolayer $MoS_2$ datasets, which are commonly used to evaluate the accuracy of E(3)-equivariant Hamiltonian frameworks, are shown in Table 1. The results for the bilayer graphene dataset are provided in the Supplementary information. Both GTP implementations share the same tensor product paths, so they are expected to achieve comparable accuracy. This is confirmed by the results of Hot-Ham. Our models demonstrate state-of-the-art accuracy across all systems, achieving 0.08, 0.12 and 0.15meV MAEs for monolayer graphene, monolayer $MoS_2$ and bilayer graphene respectively. Notably, compared to the models achieving the best performance in each system, our models require significantly fewer parameters.

## Multilayer twisted $MoS_2$

2D multilayer materials have garnered significant attention due to their unique electronic, optical, and mechanical properties, which arise from their tunable interlayer coupling under different twist angles and complex stacking configurations. The twist angles between layers can lead to a range of fascinating physical phenomena, such as unconventional superconductivity and nonlinear optical effects[37]. Additionally, the stacking configuration of each adjacent layer can be independently modulated, as the weak interlayer coupling, dominated by van der Waals interactions, allows for considerable flexibility[38]. However, fabricating multilayer materials with precise controlling twist angles is not an easy task for experiments, and performing large-scale electronic structure calculations is impracticable for DFT. In the following, we demonstrate that Hot-Ham offers a general and efficient approach to studying electronic structures by investigating multilayer twisted $MoS_2$.

Through traning on bilayer and trilayer non-twisted $MoS_2$, our models achieve an accuracy of 0.20 meV. To demonstrate the generalization ability, We applied the trained model to multilayer twisted MoS$_2$ systems, including bilayer, trilayer, and double bilayer structures. For trilayer MoS$_2$, we rotated the second layer while keeping the third layer aligned with the first. In the case of double bilayer MoS$_2$, we rotated the upper two layers, as schematized with one of the rotation angles in Fig. 3c. The MAEs are shown in Fig. 3a. Despite the training set not containing any four-layer structures, our model still predicts Hamiltonians with an accuracy comparable to those of trilayer. The MAEs reach relatively large values at 21.79° twist angle, the maximum angle typically used for modeling twist systems, but their values of 1.2-1.4meV are still sufficient to accurately describe the electronic structure. The MAEs decrease gradually as the twist angle decreases, approaching the values of non-twisted systems. This can be attributed to the smaller deviation from the training set as the twist angle decreases. In addition to the powerful generalization ability, our model also exhibits high efficiency. The wall time of DFT calculations and our model's inference are displayed in Fig. 3b. The CPU and GPU calculations are performed in one node equipped with Intel Xeon Gold 6140 processors and one NVIDIA GeForce RTX 4090 respectively. The inference time of CPU and GPU are close. The CPU inference further expands the scale to $10^4$, in a time cost that increases linearly with a slope less than DFT calculations when $N > 3000$. We can easily derive band structures by diagonalizing the predicted Hamiltonians. Fig. 3d illustrates the band structure near the Fermi level for double bilayer MoS$_2$ with a twisted angle of 5.09°. The results of our model exhibit a high degree of agreement with that obtained from DFT calculations.

**Heterostructures**

By stacking 2D materials with different properties in an incommensurate manner, heterostructures exhibit novel physical effects that are significantly distinct from the intrinsic properties of individual materials. This provides a new degree of freedom to manipulate properties and band structures, thereby enriching the properties of 2D material systems. Although the lattice mismatches introduced by incommensurability and the possible twist angles make the heterostructures rather complex and rich, our model, trained on graphene/h-BN systems for example, will show its capability to accurately describe the electronic structures.

Similarly, our model is trained on the non-twisted graphene/h-BN but with a mismatch of 1.79%[39,40] corresponding to the difference of lattice constants between graphene and h-BN. These structures are generated from three stacking configurations[41], one of which is $AA$ and another two are Bernal arrangements ($AB$ and $AB'$). Fig. 4b demonstrates the good agreement on band structure for a Bernal arrangement configuration (Fig. 4a). To test the generalization ability of Hot-Ham on unseen configurations, we generate 100 heterostructures with various mismatch and twisted angles within 500 atoms by VASPKIT package[42] for band structure calculations. As shown in Fig. 4c, the predicted band eigenvalues from -1.5 eV to 1.5 eV for all the 100 structures achieve 5.46 meV MAE, while the MAE for Hamiltonians is only 0.49 meV. This accuracy is sufficient to capture significant features, such as the relatively large band gap induced by the inversion symmetry breaking, as shown in the insert of Fig. 4b. Finally, we test our model on a larger system that contains 1022 atoms (fig. 4d). The predicted band structure (fig. 4e) and density of state (DOS, fig. 4f) agree well with results obtained from OpenMX. These again indicate that Hot-Ham can serve as a robust and powerful approach to exploring the electronic structures of a wide variety of configurations, even though trained on a limited dataset.

**Phosphorus allotropes**

In addition to 2D materials, we also evaluated Hot-Ham's performance on bulk systems using phosphorus allotropes as an example. We have collected 10 allotropes from Materials Project[43], which contains 2D and bulk systems. Among these structures, the violet phosphorus that has been experimentally synthesized recently is found to be the most stable allotrope at ambient pressure[44] and is thought as a promising semiconducting material for photonic and electronic applications. Therefore, we use the violet phosphorus to demonstrate Hot-Ham's generalization ability, while the remaining 9 allotropes are used for training.

. Our model achieves an accuracy of 0.68meV MAE for the Hamiltonian matrix on the test set. Then we verify the accuracy on violet phosphorus. The violet phosphorus shown in Fig. 5a has 84 atoms in unit cell, featuring tubular P2[P8]P2[P9] strands arranged perpendicularly. The MAE of Hamiltonian matrix is only 0.89meV, enabling our model to accurately reproduce the band structure in Fig. 5b. This example demonstrates that Hot-Ham can effectively represent features of atoms and bonds in diverse chemical environments, showcasing its potential for applications in high-throughput material discovery, such as crystal structure searches[45] and functional materials predictions.

## Discussion

In this work, we develop Hot-Ham, an E(3)-equivariant message passing neural network that combines local coordinate transformation and GTP to efficiently model DFT Hamiltonians from material structures. Through a local coordinate transformation, SO(2) convolution significantly reduces the CGTP computational cost from $O(L^6)$ to $O(L^3)$ for the convolution operations. While GTP provides a more efficient and general approach to implement tensor products, either through convolving in 2D Fourier bases with $O(L^3)$ or by spherical convolution with $O(L^2 log^2 L)$, however, at the expense of the absence of antisymmetric tensors. Our method not only retains the advantage of high efficiency of GTP, but also compensates for the antisymmetric tensors through SO(2) convolution, achieving an accurate, generalizable, and efficient E(3)-equivariant framework for electronic structure calculations for large systems. The comparison with other E(3)-ENN models demonstrate Hot-Ham's state-of-the-art accuracy with a significantly smaller parameter size compared to most models. And the experiments on MoS$_2$, heterostructures, and phosphorus allotropes show Hot-Ham's good generalization ability to structures out of training sets. In the future, it should be very valuable to expand the application scope of Hot-Ham to overcome challenges faced by traditional ab initio methods. For example, one can derive forces from predicted Hamiltonians, allowing for the explicit incorporation of electronic effects in molecular dynamics simulations. In addition, by utilizing automatic differentiation techniques, one would be able to investigate the electron-phonon interactions for large-scale systems. Furthermore, the prediction of orthogonal-base Hamiltonians to interface with the linear scaling quantum transport methodologies[46,47] is also a possible direction. In summary, our Hot-Ham method is a promising framework with great potential for large-scale electronic calculations, which is fundamentally important for designing electronic devices.

# Materials and Method

## Dataset

The Hamiltonians and overlap matrixes are all calculated by OpenMX using Perdew-Burke-Ernzerhof[48] (PBE) functional and normconserving pseudopotential. The PAO employed for B, C, N, P, S, Mo are B7.0-s2p2d1, C6.0-s2p2d1, N6.0-s2p2d1, P7.0-s2p2d1, S7.0-s2p2d1, Mo7.0-s3p2d2 respectively. We use $3 \times 3$ supercell and $6.50\text{Å}$ interlayer spacing for multilayer $MoS_2$, $4 \times 4$ supercell and $3.22\text{Å}$ interlayer spacing for graphene/h-BN, and we enlarge the phosphorus allotropes to obtain supercells with 32 to 64 atoms. In addition to applying 300K random perturbations by Phonopy package, we also performed random interlayer slides for multilayer $MoS_2$ and graphene/h-BN up to their lattice constants. For each system, the dataset is splitted for training, testing, and validation by 3:1:1. For structures used for band structure predictions, the twisted multilayer $MoS_2$ were modeled by ASE package[49]. The graphene/h-BN heterostructures with no more than 500 atoms generated by the VASPKIT package are all within 1.79% mismatch, and the 1022 atoms twisted heterostructure was also modeled by the ASE package.

## Details of neural network and training

In our model, we use the Chebyshev basis to expand the interatomic distance $|\mathbf{r}_{ij}|$:

$$RBF(|\mathbf{r}_{ij}|)_k = \frac{1}{2}\left[T_k\left(2\left(\frac{r_{ij}}{r_c}-1\right)^2\right)+1\right]f_c(|\mathbf{r}_{ij}|)$$

Where $r_c$ is the cutoff distance, $T_k$ is the $k$th order Chebyshev polynomial of the first kind, and $f_c$ is the cutoff function:

$$f_c(|\mathbf{r}_{ij}|) = \frac{1}{2}\left[1+cos\left(\pi\frac{|\mathbf{r}_{ij}|}{r_c}\right)\right]$$

The features passed to gate have the form:

$$x = \left(\oplus_{c_1} x_{1,c_1 001}\right) \oplus \left(\oplus_{c_2 l_2 m_2} x_{2,c_2 l_2 m_2 p_2}\right)$$

Then it is nonlinear as:

$$Gate(x)_{clmp} = \begin{cases} silu(x_{2,clmp}), & l = 0, p = 1 \\ tanh(x_{2,clmp}), & l = 0, p = -1 \\ sigmoid(x_{1,c001})x_{2,clmp}, & l > 0 \end{cases}$$

The target of the neural network is to minimize the following loss function:

$$L(H) = \frac{1}{N_e}\sum_{i=1}^{N}\sum_{\alpha,\beta=0}^{N_i-1} \left|H_{i,\alpha\beta} - H_{i,\alpha\beta}^{ref}\right| + \sqrt{\frac{1}{N_e}\sum_{i=1}^{N}\sum_{\alpha,\beta=0}^{N_i-1}\left(H_{i,\alpha\beta} - H_{i,\alpha\beta}^{ref}\right)^2}$$

Where $N$ is the number of structures, $N_i$ is the dimension of $i$th Hamiltonian, and $N_e$ is the total number of matrix elements. Parameters were optimized by the AdamW[50,51] algorithm with an initial learning rate of 0.02. To accelerate convergence, we use the ReduceLROnPlateau method to schedule the learning rate: the learning rate will be reduced by a factor of 0.9 if accuracy is not improved within 50 epochs. We trained 4000 epochs in graphene/h-BN, and 3000 epochs in the remaining systems. All models were trained on one NVIDIA 4090 GPU. More hyperparameters are provided in Supplementary information.

# Acknowledgments

We gratefully acknowledge the financial support from the National Natural Science Foundation of China (grant number. 12125404, T2495231, 123B2049), the Basic Research Program of Jiangsu (Grant BK20233001, BK20241253), the Jiangsu Funding Program for Excellent Postdoctoral Talent (2024ZB002, 2024ZB075), the Postdoctoral Fellowship Program of CPSF (Grant GZC20240695), the AI & AI for Science program of Nanjing University, and the Fundamental Research Funds for the Central Universities. The calculations were carried out using supercomputers at the High Performance Computing Center of Collaborative Innovation Center of Advanced Microstructures, the high-performance supercomputing center of Nanjing University.

# Figures and table

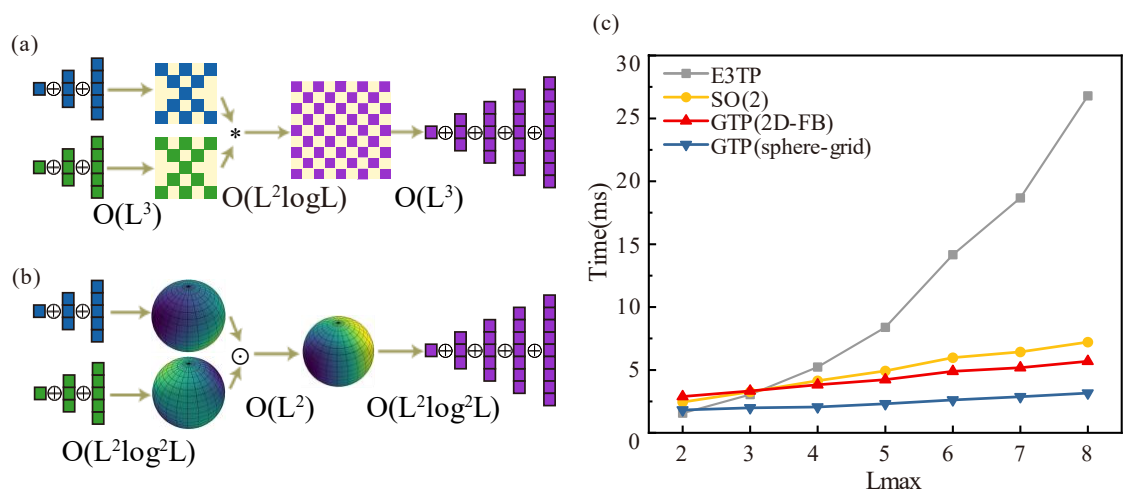

**Fig. 1. Illustration of Gaunt tensor product (GTP) method and comparison of tensor product efficiency. a** GTP (2D-FB) method. Spherical harmonics features are transformed into 2D Fourier bases representation. After FFT convolution, they are transformed back into spherical harmonic representation. **b** GTP (sphere-grid) method. Spherical harmonics features are projected into sphere grid via inverse FFT, followed by element-wise product, and finally converted back to spherical harmonic representation. **c** Comparison of efficiency for CGTP (implemented by e3nn, labeled as E3TP), SO(2) convolution, GTP (2D-FB), and GTP (sphere-grid).

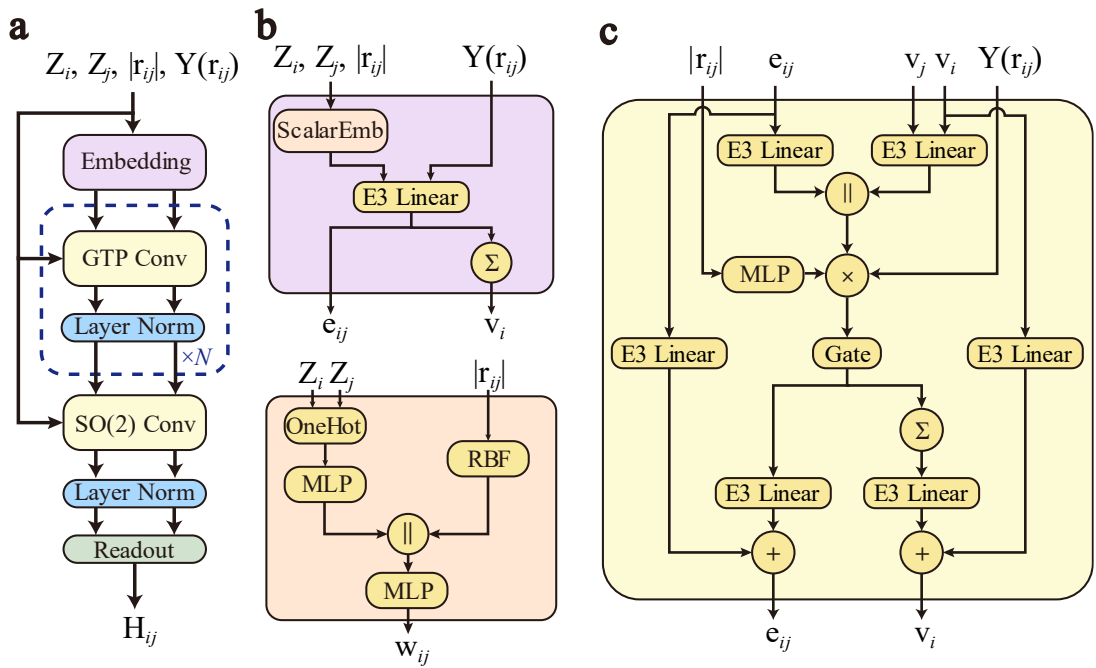

**Fig. 2. The architecture of Hot-Ham. a** The overall architecture of Hot-Ham. Atom species and relative positions are embedded into initial node and edge features. After several iterative updates in convolution layers, the final node and edge features are used to construct Hamiltonians. **b** the embedding layer. The scalar information including atom species $Z_i, Z_j$ and interatomic distance $|r_{ij}|$ is embedded through the ScalarEmd block into weight $w_{ij}$, and then is used to weight the spherical harmonics functions $Y(r_{ij})$. **c** The convolution layer. Node and edge features are concatenated to serve as messages. After the gate (activation function), messages are used to generate new node and edge features simultaneously.

**Table. 1. Comparison of the results by DeepH-E3, HamGNN, DeepH-2, DeePTB, and Hot-Ham on Hamiltonians of monolayer graphene and monolayer MoS$_2$.** (MAEs are in unit of meV. Parentheses indicate the number of parameters. For each model we only show their best performance found in previous work.)

|  | DeepH-E3[14] | HamGNN[14] | DeepH-2[13] | DeePTB[14] | Hot-Ham (2D-FB) | Hot-Ham (sphere-grid) |
|---|---|---|---|---|---|---|
| **Monolayer Graphene** | 0.28(4.5M) | 0.17(4.3M) | 0.12 | 0.14(4.5M) | **0.08(0.9M)** | **0.07(0.9M)** |
| **Monolayer MoS$_2$** | 0.46(1.0M) | 0.37(4.3M) | 0.21 | 0.14(4.5M) | **0.12(1.9M)** | **0.12(1.9M)** |

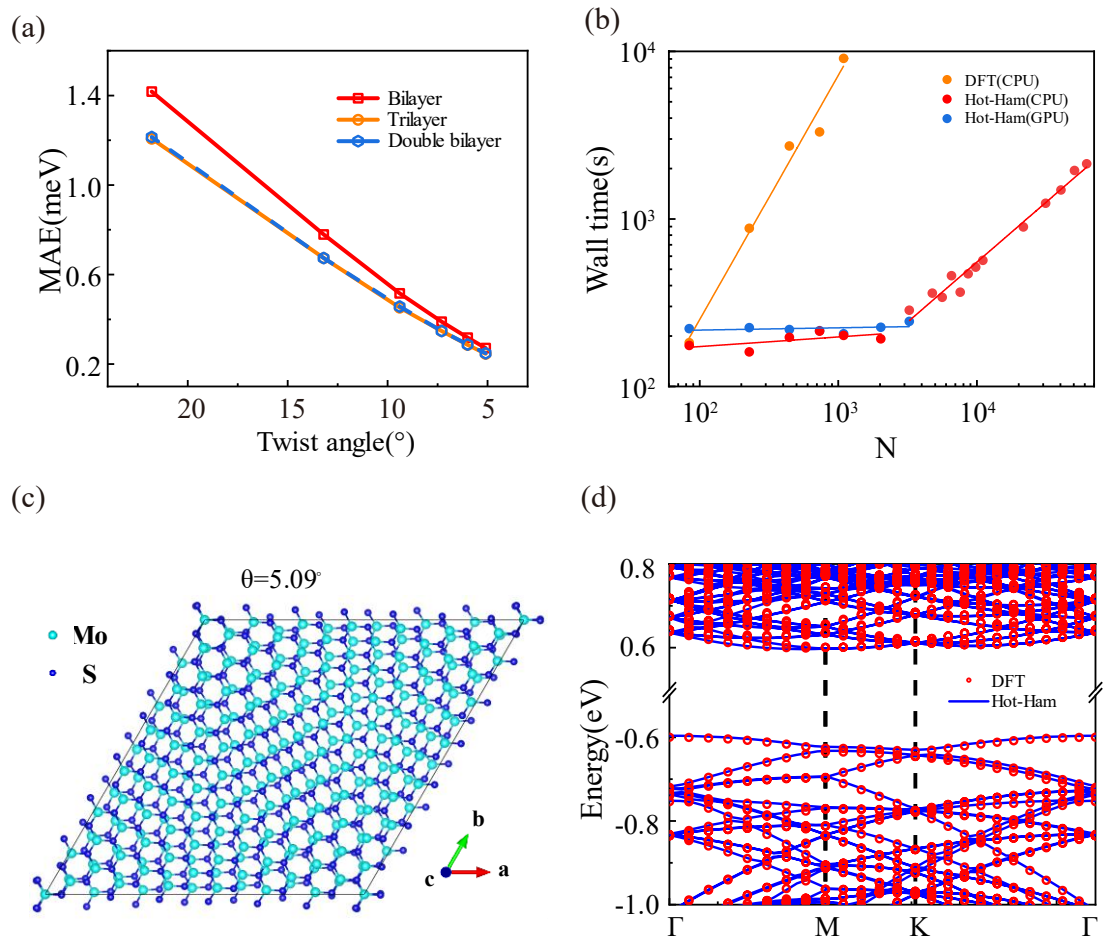

**Fig. 3. Application of Hot-Ham to multilayer MoS$_2$. a** Hamiltonian MAE for bilayer, trilayer, and double bilayer under different twist angles. **b** The wall time for DFT calculations and our model's inference at CPU and GPU. **c** structure of double bilayer MoS$_2$ containing 1524 atoms with twist angle $\theta = 5.09°$. **d** band structure for the structure in **c** predicted by DFT (OpenMX) and Hot-Ham.

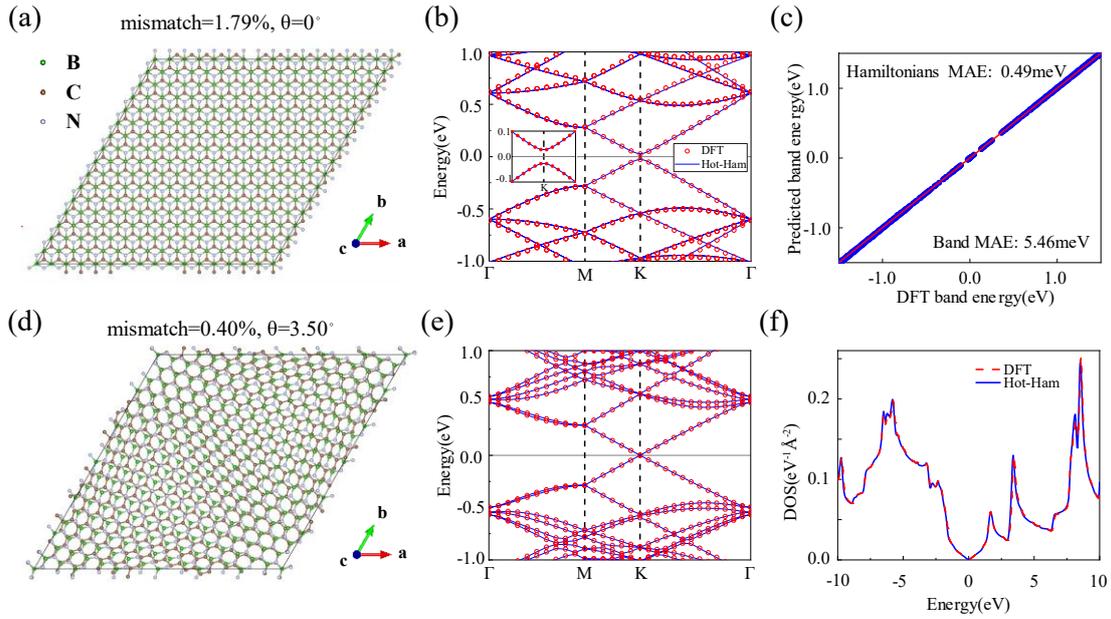

**Fig. 4. Performance of Hot-Ham on graphene/h-BN heterostructures. a** $16 \times 16$ supercell cell of a Bernal arrangement graphene/h-BN heterostructure. **b** the band structure of **a**. **c** comparison of band eigenvalues within -1.5~1.5 eV for the 100 heterostructures with various mismatch and twisted angles. The MAE of their Hamiltonians is also displayed. **d** structure of graphene/h-BN containing 1022 atoms (518 C atoms, 252 B/N atoms) with mismatch=0.4% and twist angle $\theta = 3.50°$. **e** and **f** are band structure and DOS for structure **d** predicted by DFT (OpenMX) and Hot-Ham.

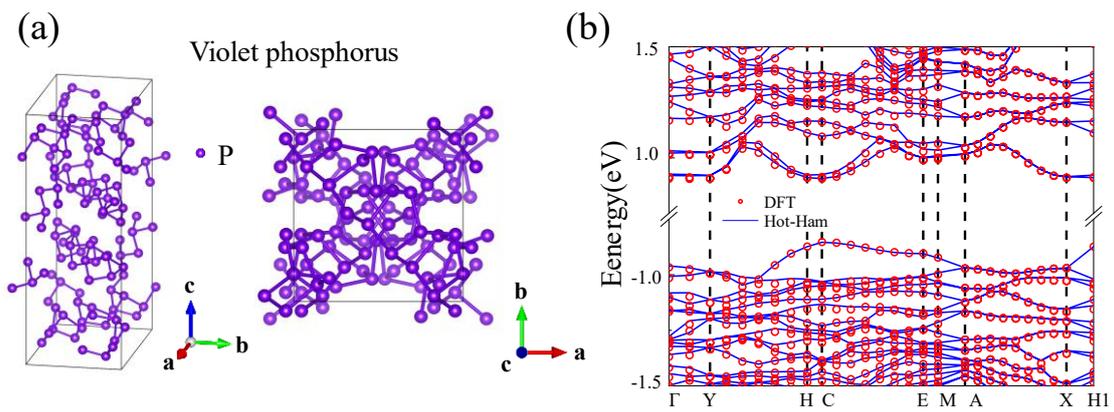

**Fig. 5. Predicted electronic structure on phosphorus allotropes. a** crystal structure of violet phosphorus. **b** band structure calculated by DFT (OpenMX) and Hot-Ham for the structure that is not in the training dataset.